\documentclass{bioinfo}
\copyrightyear{2005}
\pubyear{2005}

\usepackage{bm,amsmath,amssymb,graphicx,subfig,multirow}

\begin{document}
\firstpage{1}

\title[Network Inference Using Goldbeter-Koshland Kinetics]{Network Inference Using Steady-State Data and Goldbeter-Koshland Kinetics}
\author[Oates and Mukherjee]{Chris J Oates\,$^{1,2,3}$,
Bryan T Hennessy$^{4}$, Yiling Lu$^{5}$, Gordon B Mills$^{5}$ 
and Sach Mukherjee\,$^{3}$}
\address{$^{1}$Centre for Complexity Science, University of Warwick, CV4 7AL, UK.\\
$^{2}$Department of Statistics, University of Warwick, CV4 7AL, UK.\\
$^{3}$Department of Biochemistry, Netherlands Cancer Institute, 1066CX Amsterdam, The Netherlands.\\
$^{4}$Department of Medical Oncology, Beaumont Hospital, Dublin, Ireland.\\
$^{5}$Department of Systems Biology, The University of Texas M. D. Anderson Cancer Center, Houston, TX 77030, USA.}

\history{Received on XXXXX; revised on XXXXX; accepted on XXXXX}

\editor{Associate Editor: XXXXXXX}

\maketitle

\begin{abstract}

\section{Motivation:}
Network inference approaches are widely used to shed light on regulatory interplay between molecular players such as genes and proteins.
Biochemical processes underlying networks of interest (e.g. gene regulatory or protein signalling networks) are generally nonlinear. In many settings, knowledge is available concerning relevant chemical kinetics. However, existing network inference methods for continuous data are typically rooted in convenient statistical formulations which do not exploit chemical kinetics to guide inference. 
\section{Results:}
Here we present an approach to network inference for steady-state data that is rooted in nonlinear descriptions of biochemical mechanism. 
We use equilibrium analysis of chemical kinetics to obtain functional forms that are in turn used to infer networks using steady-state data. 
The approach we propose is directly applicable to conventional steady-state gene expression or proteomic data and does not require knowledge of either network topology or any kinetic parameters; both are simultaneously learned from data.
We illustrate the approach in the context of  protein phosphorylation networks, using data simulated from a recent mechanistic model and proteomic data from cancer cell lines. In the former, the true network is known and used for assessment, whilst in the latter results are compared against known biochemistry. We find that the proposed methodology is more effective at estimating network topology than  methods based on linear models. 
\section{Availability:}
MATLAB R2009b code used to produce these results is provided in the Supplemental Information.

\section{Contact:} \href{c.j.oates@warwick.ac.uk}{c.j.oates@warwick.ac.uk}, \href{s.mukherjee@nki.nl}{s.mukherjee@nki.nl}
\end{abstract}

\section{Introduction}

Networks of molecular components play a prominent role in molecular and systems biology. 
A graph $G = (V,E)$ can be used to describe a biological network, with vertex set $V(G)$ identified with molecular components (e.g. genes or proteins) and edge set $E(G)$ with 
regulatory interplay between the components. Edges in a biological network are often associated with the causal notion
that intervention on a parent node influences its child node(s).
Data-driven characterisation of the graph structure $E(G)$ 
(often referred to as the {\it topology}) 
is known as {\it network inference} and has emerged as an important problem class in bioinformatics and systems biology \citep{Chou}.
Network inference can aid in efficient generation of biological hypotheses from high-throughput data.
Further, network inference can aid in exploring molecular interplay that is associated with specific phenotypes, such as disease states. 

From a statistical perspective, network inference entails reverse-engineering a graph $G$ using 
biochemical data $\mathcal{D}$ and, where available, prior knowledge regarding aspects of the topology.
Over the last decade 
many methods for network inference have been proposed, with popular approaches are reviewed in \cite{Bansal,Hecker,Lee,Markowetz}.
To date, most methods for network inference have been rooted in discrete or linear 
formulations \citep{Bender,Sachs,Morrissey,Opgen,Hill,Kim}. 
As discussed in \citet{Oates}, a wide range of existing approaches can be viewed as variants of the linear regression model in statistics.
(In this paper ``linear'' refers to linearity in parameters, so that nonlinear basis functions may be used within a ``linear'' framework.)
Moreover, a number of approaches based on ordinary differential equations (ODEs; \cite{Bansal,Li,Nam}) are ultimately reducible to linear statistical models, as described in \cite{Oates3}. 

However, the biochemical processes underlying biological networks are often highly nonlinear. 
When the data-generating process is nonlinear, use of linear models may produce inefficient or inconsistent estimation, attributing causal status to artifacts resulting from model misspecification \citep{Heagerty,Lv}.
Indeed, such bias can prevent recovery of the correct network even in favourable asymptotic limits of large sample size and low noise \citep{Oates}.
On the other hand, in many settings nonlinear dynamical models of  relevant biochemical processes are available. 
For example gene regulation may be modelled using Michaelis-Menten functionals \citep{Cantone}, and metabolism may be modelled using mass action chemical kinetics \citep{Min}.
Here, we describe an approach by which kinetic models can be used to inform network inference from steady-state data.
As we show below, such information can be valuable in guiding exploration of network topologies.

Kinetic formulations have been widely studied in the systems biology literature \citep{Bintu} and recently there has been much interest in statistical inference for such systems, with examples including \cite{Chen,Xu}. Our work is in a similar vein, but focuses on network inference {\it per se} and on the steady-state rather than time-course setting.
While biochemical assays have become cheaper, it remains the case that experimental designs must often negotiate a trade off between more conditions (e.g. perturbations, biological samples, technical replicates) and temporal resolution (e.g. number of time points). Methodologies which can exploit knowledge concerning relevant dynamical systems in the steady-state setting are therefore potentially valuable. 

In brief, we proceed as follows. 
We consider a class of nonlinear biochemical dynamical systems that are relevant to the biological process of interest (we focus on protein signalling, discussed in detail below).
Steady-state analysis leads to a class of functional relationships 
between parent and child. These functional relationships are used to formulate a statistical model for network inference from steady-state data. 
In this way, network inference is rooted in functional relationships derived from nonlinear kinetics. 
Importantly, we do not assume detailed knowledge of the dynamical system, but only the broad class to which dynamics and associated equilibria may belong. 
Indeed, the approach we describe does not require any kinetic parameters to be known {\it a priori}, nor knowledge of the network topology, and is
in that sense directly comparable to conventional network inference methods. 
Its potential advantage stems from then rich yet constrained nature of the class of functional relationships that are considered. 
As recently discussed in \citet{Peters},
nonlinear functional forms can aid in identification of underlying causal relationships.

We develop these ideas in the context of protein signalling mediated by phosphorylation.
Enzyme kinetics have been extensively studied, and dynamical formulations are widely available in the literature \citep[see e.g.][]{Leskovac}. For some proteins and pathways, regulation has been studied in considerable causal and mechanistic detail.
Indeed, there exist detailed computational models for canonical protein signalling pathways, which have been  validated against experimental data \citep[e.g.][]{Schoeberl,Xu}. Further, proteomic technologies now allow multivariate, data-driven study of phosphorylation, facilitating biological validation of proposed methodology. We take advantage of these factors to examine the performance of our methodology using both simulated and real data.

In the phosphorylation setting, Goldbeter-Koshland kinetics \citep{Goldbeter} form the functional class that underlies our network inference approach. Goldbeter-Koshland kinetics are well known to be capable of highly nonlinear behaviour including exquisite sensitivity. 
It has been experimentally demonstrated that this so-called {\it ultrasensitivity} is biologically relevant to signalling network dynamics, facilitating abrupt and precise decision making (e.g. \cite{Kim2}).
We carry out statistical inference in a Bayesian framework, using reversible-jump Markov chain Monte Carlo (RJMCMC) to explore the joint model and parameter space.
This yields posterior probability scores for edges in the network that are analogous to scores obtained in existing statistical network inference approaches for steady-state data \cite{Hill}.

The remainder of this paper is organised as follows. In Section \ref{methods} our approach is laid out, followed by a detailed exposition of the associated computational statistics. In Section \ref{results} we present results on data simulated from a recently developed dynamical model of the mitogen-activated protein kinase (MAPK) signalling, that has been validated against experimental data \citep{Xu}. We then show results on real proteomic data from breast cancer cell lines.
Finally, Section \ref{discussion} closes with a discussion of practical implications and opportunities for network inference based on functional models, along with associated technical challenges.

\begin{methods}
\section{Methods} \label{methods}

\begin{figure*}[!tpb]
\centering
\subfloat[]{\includegraphics[width = 0.28\textwidth,clip,trim = 7.3cm 18.5cm 7.2cm 4.4cm]{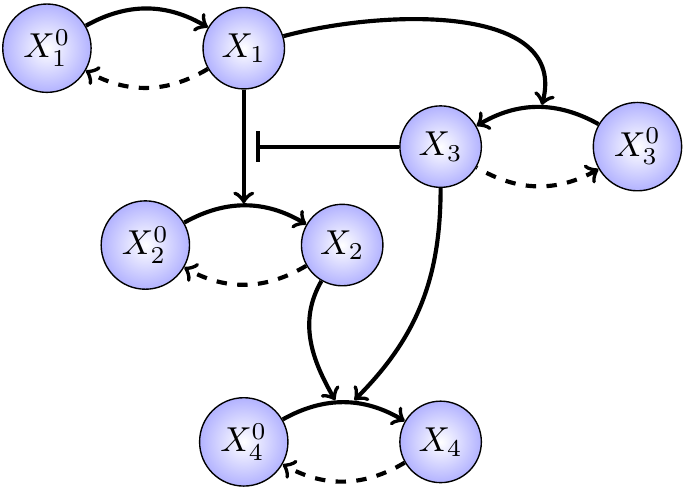}\label{example}} \; \; \;
\subfloat[]{\includegraphics[width = 0.28\textwidth,clip,trim = 7.7cm 19cm 7.7cm 4.4cm]{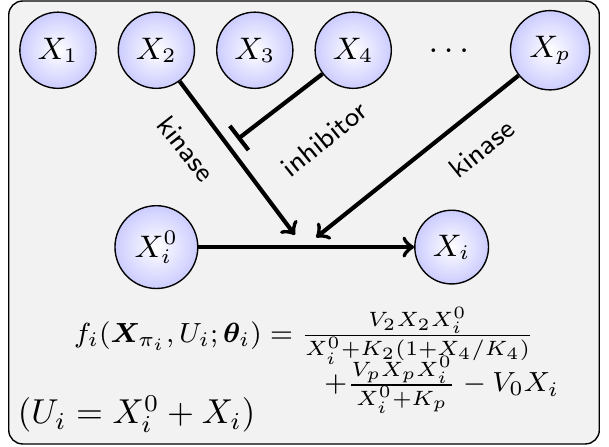}\label{fig: plan}} \; \; \;
\subfloat[]{\includegraphics[width = 0.28\textwidth,clip,trim = 3cm 22.9cm 12cm 2.5cm]{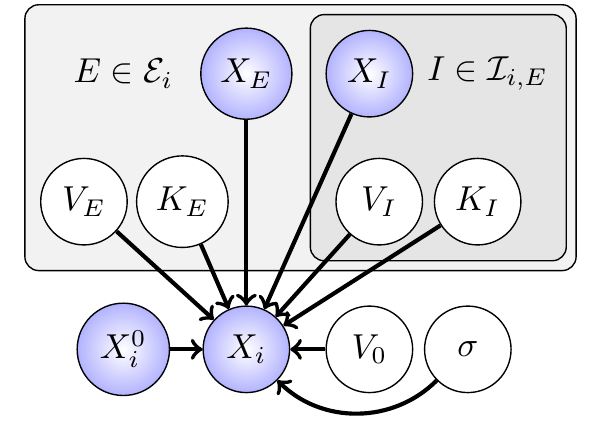}\label{graphical model}}
\caption{Overview of approach. (a) An example of a phosphorylation network. (b) Our approach couples automatic generation of chemical models with Bayesian model selection to infer regulators $\pi_i$ of species $i$. (c) A statistical formulation (graphical model) for equilibrium phosphorylation of species $i$ is characterised by specifying kinases ($E \in \mathcal{E}_i$) and inhibitors ($I \in \mathcal{I}_{i,E}$) of kinases. [Bounding boxes in this case are used to indicate multiplicity of variables, shaded nodes are observed with noise.]}
\end{figure*}

We begin in Section \ref{generalmodel} by describing our approach in general terms. Section \ref{chemistry} then introduces relevant concepts in the application area of protein phosphorylation. 
In particular we describe a class of nonlinear equations derived from Goldbeter-Koshland kinetics. 
Next, in Section \ref{statistics} this model class is embedded into a Bayesian statistical framework for observations obtained at equilibrium.
Inference over model space is facilitated by RJMCMC, with Section \ref{RJMCMC} dedicated to a presentation of our sampling scheme and a discussion of key implementational details.

\subsection{General Formulation} \label{generalmodel}

We consider a state vector $\bm{X} = (X_1,\dots,X_p)$ containing concentrations of $p$ proteins. 
Equilibrium analysis of phosphorylation dynamics, as described below, leads to a system of $p$ equations $X_i = f_i(\bm{X},U_i;\bm{\theta}_i)$ where $i$ indexes proteins, $U_i$ are external input variables and $\bm{\theta}_i$ unknown parameters. 
The component function $f_i$  depends on a subset $\pi_i$ of the state variables, such that we may write $X_i = f_i(\bm{X}_{\pi_i},U_i;\bm{\theta}_i)$, where $\bm{X}_{\pi_i}$ indicates selection of components of the vector $\bm{X}$ whose indices are members of the set $\pi_i$.
Variables $j \in \pi_i$ are the parents of node $i$ in graph $G$; 
the parent sets $\pi_i$ specify the (unknown) topology of interest
since $(j,i) \in E(G) \iff j \in \pi_i$. Our inference scheme seeks to infer the $\pi_i$'s from steady-state data.
Since the dynamical system is not usually known in  detail {\it a priori},
we consider the practically applicable case in which the $f_i$'s are known only to belong to a certain class $\mathcal{F}$ (derived from Goldbeter-Koshland kinetics, as described below) 
with  parent sets $\pi_i$ and all parameters $\bm{\theta}_i$ remaining unknown.

\subsection{Protein Phosphorylation} \label{chemistry}

We consider proteins $i \in V = \{1,\dots,p\}$, each of which has an unphosphorylated form $X_i^0$ and a phosphorylated form $X_i$ ($i \in V$).
Phosphorylated proteins are referred to as {\it phosphoproteins}.
The chemical reaction that gives product $X_i$ from substrate $X_i^0$ is known as {\it phosphorylation} and is catalysed by {\it kinases} $X_E$ ($E \in \mathcal{E}_i$). 
We consider the case in which the kinases themselves are phosphoproteins (if phosphorylation is not driven by a kinase in $V$, we set $\mathcal{E}_i = \emptyset$).
The ability of a kinase $E \in \mathcal{E}_i$ to catalyse phosphorylation of $X_i$ may be tempered by {\it inhibitors} $X_I$ ($I \in \mathcal{I}_{i,E} \subset V$; the double subscript indicates that inhibition is specific to both substrate and kinase).
Thus the parents $\pi_i$ of $X_i$ comprise both the kinases and their inhibitors: $\pi_i = \mathcal{E}_i \cup \{ \mathcal{I}_{i,E} \}_{E \in \mathcal{E}_i}$. 
Due to specificity of phosphorylation reactions, the underlying network $G$ is typically sparse, such that the number of parents $\pi_i$ for variate $X_i$ is usually low.
An example is shown, using a standard graphical representation, in Fig. \ref{example}.
In what follows we use $X_i^0,X_i$ to denote the concentrations of proteins $X_i^0,X_i$ respectively;  $U_i = X_i^0 + X_i$ is then the total concentration of protein $i$, which is taken to be approximately invariant over the timescale of phosphorylation dynamics.

For network inference, model selection will take place over parent sets $\pi_i$. 
Accordingly, we require functional equations for any such subset (Fig. \ref{fig: plan}). 
Following the  biochemical literature \citep{Kholodenko,Steijaert}, we use ODEs of the Michaelis-Menten type to provide a suitable class of analytic approximations for phosphorylation dynamics.
The rate of phosphorylation $X_i^0 \rightarrow X_i$ due to kinase $X_j$ is given by $V X_jX_i^0/(X_i^0+K)$, which explicitly acknowledges variation of kinase concentration $X_j$ and permits kinase-specific response profiles (parameterised by $K$) with maximum reaction rate $V$.

Equilibrium analysis of the foregoing kinetic model yields functional relationships between nodes that we use to inform analysis of steady-state data.
The famous example of \cite{Goldbeter} considered phosphorylation by a single enzyme ($X_E$) and dephosphorylation by a single phosphatase ($X_P$), which at equilibrium satisfy the balance equation
\begin{eqnarray}
\frac{V_EX_EX_i^0}{X_i^0+K_E} = \frac{V_PX_PX_i}{X_i+K_P}
\end{eqnarray}
whose solution $X_i = f_i((X_E,X_P),X_i^0;\bm{\theta}_i)$ is capable of expressing a range of biologically relevant nonlinearities.
In this work we extend the class of molecular regulatory mechanisms by entertaining multiple (independent) kinases along with competitive inhibition, where substrate ($X_i^0$) and inhibitor ($X_I$) compete for the same binding site on the enzyme ($X_E$):
\begin{eqnarray}
X_E X_I \rightleftharpoons  X_E \rightleftharpoons X_E X_i^0 \rightarrow X_E + X_i
\end{eqnarray}
When multiple inhibitors ($I,I'$) are present, they are assumed to act exclusively, competing for the same binding site on the enzyme:
\begin{eqnarray}
X_E X_{I'} \rightleftharpoons X_E \rightleftharpoons X_E X_{I'}
\end{eqnarray}
Mathematically, competitive inhibition by exclusive inhibitors corresponds to rescaling of the Michaelis-Menten parameter
\begin{eqnarray}
K_E \mapsto K_E \Biggl( 1 + \sum_{I \in \mathcal{I}_{i,E}} \frac{X_I}{K_I} \Biggr). 
\end{eqnarray}
where the sum ranges over inhibitors $I$ of the kinase $E$.
Phosphatase specificity is currently poorly characterised compared with kinase specificity, so our analysis does not attempt to cover this level of regulation. In particular dephosphorylation is assumed to occur at a rate $V_0X_i$ proportional to the amount of phosphoprotein.
Collecting together our modelling assumptions and solving the resulting balance equation produces a functional model class $\mathcal{F}$, with member functions $f_i \in \mathcal{F}$ given by
\begin{align}
f_i(\bm{X}_{\pi_i},U_i;\bm{\theta}_i) = \sum_{E \in \mathcal{E}_i}{ \frac{V_E/V_0 X_EX_i^0}{X_i^0+K_E\bigl(1+\sum_{I \in \mathcal{I}_{i,E}} \frac{X_I}{K_I}\bigr)}}.
\label{eq:GK}
\end{align}
Here the parameter vector $\bm{\theta}_i$ contains the maximum rates ($\bm{V}$) and Michaelis-Menten constants ($\bm{K}$) specific to phosphorylation of species $i$ (dependence of $\bm{V},\bm{K}$ on $i$ is notationally suppressed for clarity).
When $\mathcal{E}_i = \emptyset$ we instead define $f_i = \mu_i$, equal to the average phosphoprotein concentration.

\subsection{Statistical Formulation} \label{statistics}
The Goldbeter-Koshland model (\ref{eq:GK}) gives a general form for the functional relationship between nodes at steady-state.
Inference proceeds based on a Bayesian formulation of this model (Fig. \ref{graphical model}). 
Consider independent observations of protein expression obtained at equilibrium with respect to phosphorylation dynamics.
To fix a characteristic scale, all data are scale-normalised prior to inference such that each species has unit mean.
For a given protein $i$, a model $M_i$ for phosphorylation describes putative kinases $\mathcal{E}_i$ and associated inhibitors $\mathcal{I}_{i,E}$ ($E \in \mathcal{E}_i$) for protein $i$ (note that $M_i$ contains more information than the subset $\pi_i$, namely the specific mechanistic roles played by each variable in $\pi_i$). 
Then, conditional on $M_i$ and parameters $\bm{\theta}_i$ we have the following statistical model
\begin{align}
\log(X_i) = \log(f_i(\bm{X}_{\pi_i},U_i;\bm{\theta}_i)) + \epsilon_i
\end{align}
where $\epsilon_i \sim N(0,\sigma_i^2)$.
The logarithm of both predictor and response is taken in order to improve the normality assumption on the error $\epsilon_i$.

In the Bayesian setting, prior probability distributions are required for parameters $\bm{\theta}_i$ and models $M_i$. 
For the parameters $\bm{\theta}_i = (\bm{V},\bm{K},\sigma)$, which we have augmented with $\sigma$ (as with the other parameters, we drop the subscript $i$ on $\sigma$ for clarity), physical considerations require that $V_j,K_j,\sigma>0$.
Following \cite{Xu} we postulate that all biological processes must occur on an observable timescale, motivating, in the shape, scale parametrisation, the gamma priors $V \sim \Gamma(2,1/2)$, $K \sim \Gamma(2,1/2)$, each of unit mean and variance $1/2$. 
The noise parameter $\sigma$ is inverse-gamma distributed {\it a priori} as $\sigma \sim \Gamma^{-1}(6,1)$, with prior mean $1/5$ and variance $1/100$ chosen to correspond to the magnitude of measurement noise  in current proteomic technologies \citep{Hennessey}.

When expert opinion is available, rich subjective model priors may be elicited \citep[see e.g., for graphical models, ][]{Mukherjee}, but for this work we employed an objective prior, depending on a (possibly empty) prior model $M_i^0$.
Prior specification should account for the distinct roles of kinases and inhibitors; a mathematical formulation for the objective model prior is described in the Supplemental Information.

\subsection{Reversible Jump Markov Chain Monte Carlo} \label{RJMCMC}

The dimensionality of the parameter vector $\bm{\theta}^M$  depends on the model $M$; $\dim(\bm{\theta}^M) = \dim(\bm{V}^M) + \dim(\bm{K}^M) + 1$ where the former quantities are functions of the numbers of kinases and inhibitors according to $M$.
Since we seek models with high posterior probability, and noting that most models will provide insufficient explanatory power, we implement RJMCMC \citep{Green} to reduce the effective size of model space.
Following \cite{Green2} we enumerate all possible models as $\{M^{(k)}\}_{k \in \mathcal{K}}$ and define the {\it across-model} state space
\begin{eqnarray}
{\mathcal S} = \bigcup_{k \in {\mathcal K}} (\{k\} \times \Theta_k), \; \; \; 
k = \bigotimes_{E \in \mathcal{E}^{M^{(k)}}} (\{E\} \times \mathcal{I}_E^{M^{(k)}}) \label{model}
\end{eqnarray}
where parameters $\bm{\theta}^{M^{(k)}}$ for model $M^{(k)}$ belong to $\Theta_k$ and $\otimes$ denotes the Cartesian product.
The reversible jump sampler constructs an ergodic Markov chain on ${\mathcal S}$ which has, as its stationary distribution, the posterior probability distribution $p(s|\mathcal{D}), s \in \mathcal{S}$. In particular the marginal  $p(k|\mathcal{D})$ over the model index $k \in {\mathcal K}$ corresponds exactly to the posterior model probabilities $p(M^{(k)}|\mathcal{D})$.
Construction of an efficient RJMCMC sampler requires an intuition for the across model state space. We  adopt a deliberately transparent Metropolis-within-Gibbs approach \citep{Roberts2}, updating one coordinate of ${\mathcal S}$ at a time using a Metropolis-Hastings accept/reject probability of the form $\alpha(s,s') = \min(1,A(s,s')p(\mathcal{D}|s')/p(\mathcal{D}|s))$.
A number of distinct proposal mechanisms were employed in order to ensure ergodicity and provide rapid mixing. Precise details of the proposals used, along with their associated ratios $A(s,s')$ may be found in the Supplemental Information.
For applications, 30,000 iterations of the Gibbs sampler were performed, with 5,000 discarded as burn-in.
Convergence was assessed using repeated runs from dispersed initial conditions.

\end{methods}

\section{Results} \label{results}

In this Section we empirically assess our methodology and compare its performance against network inference based on the linear model.
In Section \ref{silico} we show results using a recently published dynamical model of the MAPK signalling pathway due to \cite{Xu}, where the underlying network is known exactly. 
In Section \ref{vitro} we apply our approach to a real proteomic dataset, which has an unknown and presumably more complex noise structure.
In both cases, for fair comparison between different methods, no informative model priors were used (i.e. we set $\forall i, M_i^0 = \emptyset$).

\subsection{Simulation Study} \label{silico}

\begin{figure*}[!ht]
\centering
\subfloat[]{\includegraphics[width = 0.33\textwidth,clip,trim = 4.3cm 12cm 3cm 4.4cm]{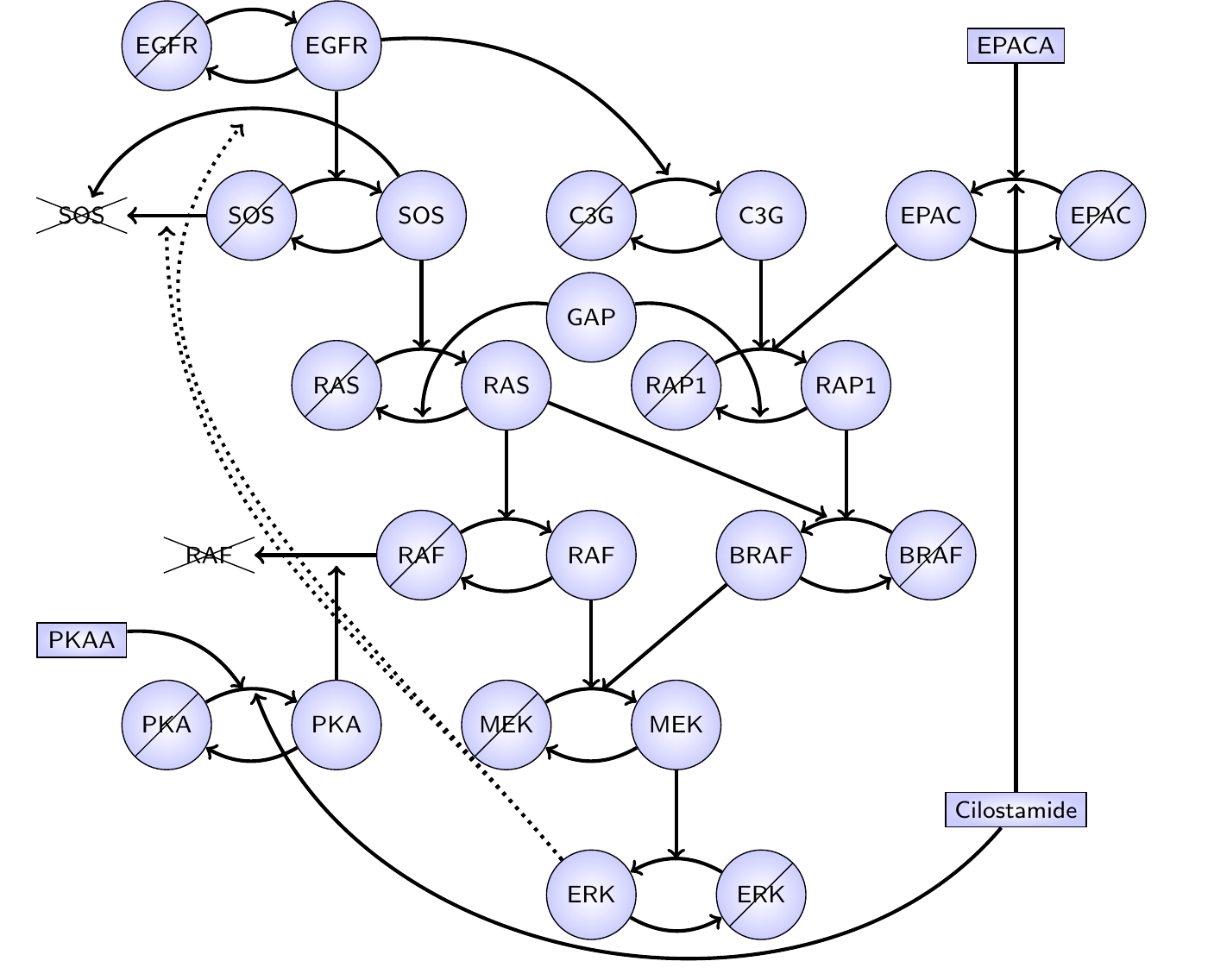}\label{fig: Xu}}
\subfloat[]{\includegraphics[width = 0.33\textwidth,clip,trim = 3cm 8.8cm 3.5cm 8.8cm]{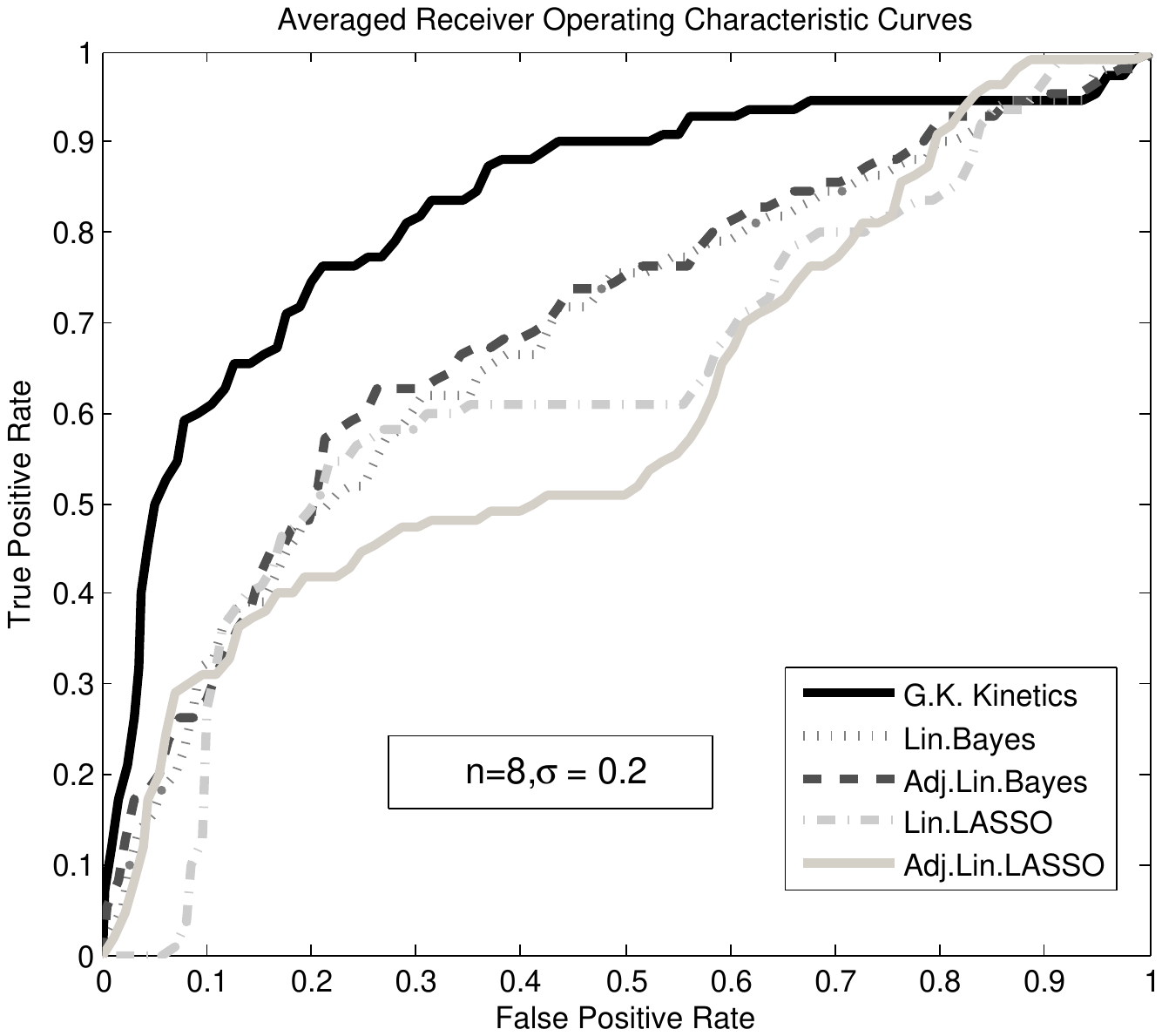}\label{ROC}}
\subfloat[]{\includegraphics[width = 0.3\textwidth,clip,trim = 4.5cm 8.8cm 4cm 9cm]{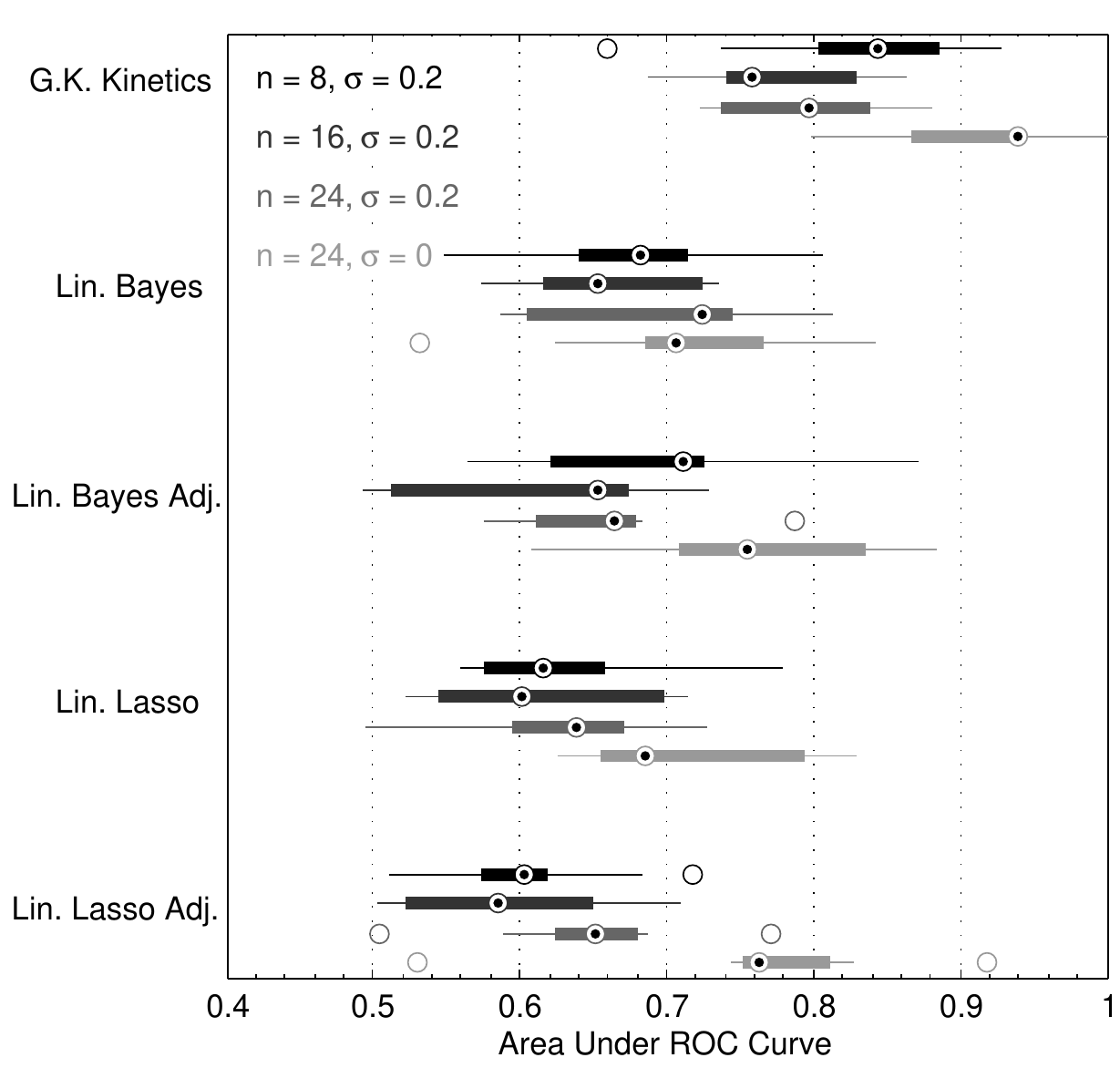}\label{AUR}}
\caption{Simulation study. (a) Computational model of the MAPK signalling pathway \citep[due to][]{Xu}. Circles represent proteins, rectangles represent interventions (drug treatments) used to perturb the system. For proteins, one strike-through represents inactivity, two strikes represent degradation. (b) Average receiver operating characteristic (ROC) curves (sample size $n = 24$, noise $\sigma = 0.2$, see text for details) using data generated from model  (a). (c) Area under ROC curve (AUR) for each of the sample size ($n$) and noise ($\sigma$) regimes shown (boxplots over 10 datasets for each $n, \sigma$ regime). [Legend: ``G.K. Kinetics": network inference using Goldbeter-Koshland kinetics as described in text; ``Lin. Bayes": Bayesian variable selection using linear model; ``Lin. Lasso": variable selection using LASSO and linear model; ``Lin. Bayes Adj."  and ``Lin. Lasso Adj.": as previous but corrected for total protein levels as described in text.]}
\end{figure*}

Data were generated from a computational model of the MAPK signaling pathway due to  \cite{Xu}, specified by a system of 25 nonlinear ODEs (Fig. \ref{fig: Xu}).  
The simulation gives covariates that are highly correlated at equilibrium, as would be expected in practice, whilst providing a known network $G$ for evaluation purposes.
Further details regarding the computational model are described in the Supplemental Information.
We introduced independent Gaussian measurement noise, additive on the log scale, of magnitude $\sigma = 0.2$, similar to technical error incurred by current proteomic technologies \citep{Hennessey}.

We benchmarked our approach against the linear-additive-Gaussian formulation $\log(\bm{X}_i) \sim N(\bm{1}\beta_0+\bm{D}_M\bm{\beta}_M,\sigma^2\bm{I})$ with design matrix $\bm{D}_M = [\dots \log(\bm{X}_j) \dots]_{j \in \pi^M}$ and intercept $\beta_0$; the logarithm of a vector is taken component-wise. 
All variables were mean-variance standardised prior to inference.
We consider two standard approaches to inference for the linear model, namely (1) the LASSO with penalty parameter set according to cross validation (``Lin. Lasso"), and (2) a conjugate Bayesian formulation (``Lin. Bayes"; \cite{Hill}), based on the $g$-prior $\bm{\beta}_M \sim N(\bm{0},n\sigma^2(\bm{D}_M'\bm{D}_M)^{-1})$, with a flat prior over the intercept $p(\beta_0) \propto 1$ and reference prior over the noise $p(\sigma) \propto 1/\sigma$. For the Bayesian approach we took a model prior $p(M)$ to be uniform over in-degree $d = \dim(\bm{\beta}_M)$ with the restriction $d \leq 3$. 
Model averaging was then used to obtain posterior inclusion probabilities.
For each of the linear approaches (1) and (2) we also considered {\it adjusted} variants 
(``Lin. Lasso Adj." and ``Lin. Bayes Adj.") where log-phospho-ratios $\log(X_i/U_i)$ constitute the response; this can be motivated as a simple first order correction for variation in total protein levels.

For each phosphorylated or active species $i$ in the computational model, we sought to infer the parents $\pi_i$. For a fair comparison with the linear approaches, which do not ascribe functional roles to variables, we did not distinguish between kinases and  inhibitors during assessment.
The resulting receiver operating characteristic (ROC) curves are shown in Fig. \ref{ROC}.
Overall performance was quantified using area under the ROC curve (AUR),
aggregated  over all $i \in V$.
Results are shown over 10 datasets $\mathcal{D}$ for each of various combinations of sample size $n$ and noise level $\sigma$ (Fig. \ref{AUR}).
In all regimes our approach outperformed linear approaches; the latter did not perform well even in this low dimensional example.
We note also that even in the least challenging regime ($n = 24$, $\sigma = 0$), none of the approaches were able to perfectly recover the entire network $G$.
The adjusted regressions, which model the log-phospho-ratio as the response, did not outperform the standard linear regressions.

\subsection{Cancer Proteomic Data} \label{vitro}

\begin{figure}[!h]
\centering
\subfloat[Phosphoproteomic data obtained from breast cancer cell lines]{\includegraphics[width = 0.5\textwidth,clip,trim = 2cm 6cm 2cm 6cm]{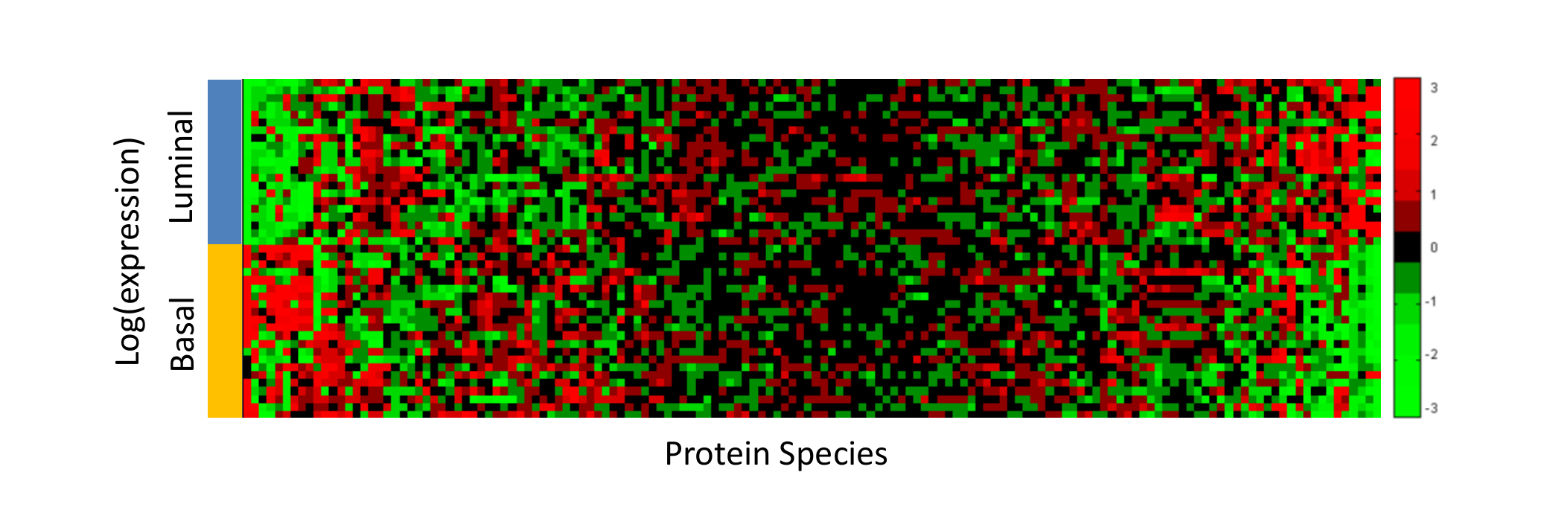}\label{VitroData}}

\subfloat[Inferred parents (ranked) for S6, based on basal cell line data]{\includegraphics[width = 0.4\textwidth,clip,trim = 3.5cm 9cm 3.9cm 9cm]{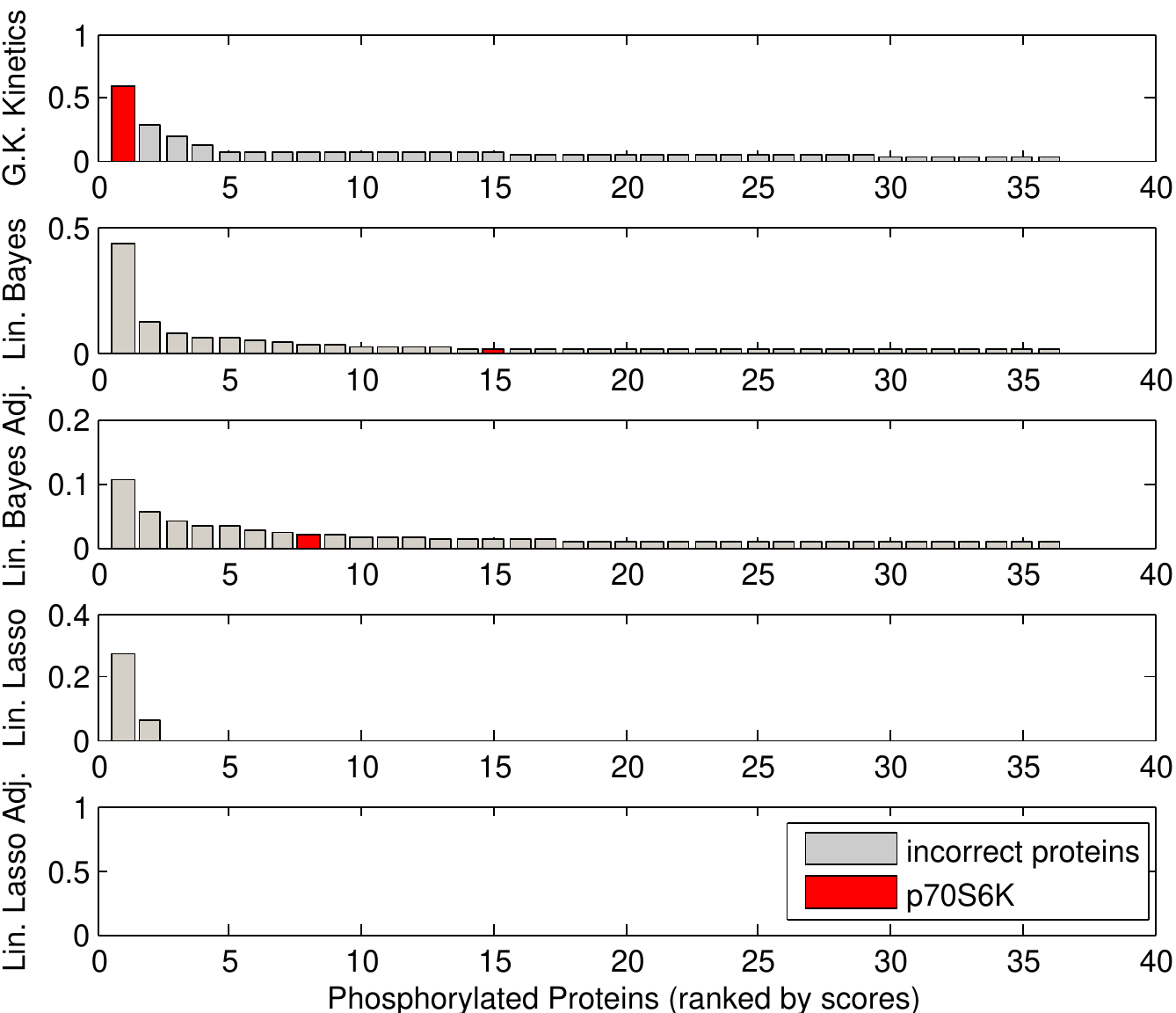}\label{ranks}}

\subfloat[Comparison of competing methodologies]{\includegraphics[width = 0.4\textwidth,clip,trim = 1cm 0.5cm 1cm 0.5cm]{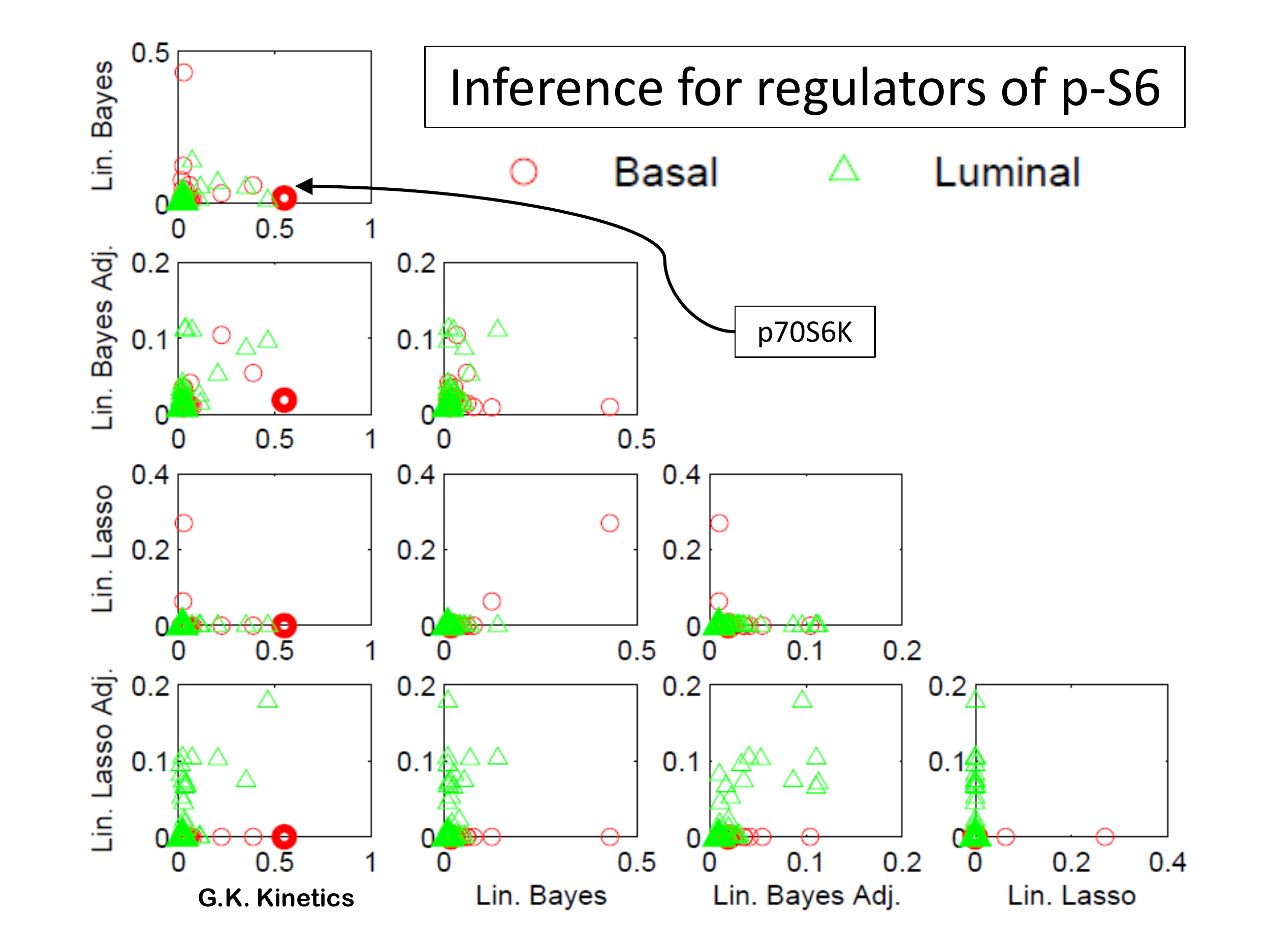}\label{Vitro}}
\caption{Cancer protein data. (a) Heatmap of reverse-phase protein array data from a panel of breast cancer cell lines. (b) Proteins were ranked as potential regulators of the node S6 under each methodology. The protein p70 S6 Kinase (p70S6K) is known to be a key kinase for the node S6; this known regulator is shown in red in the bar plots. (c) Comparison of  methodologies: Each point in the scatter plots represents one phosphoprotein, with the known kinase p70S6K highlighted in bold. [(b) and (c) display weights (posterior probabilities or absolute regression coefficients) assigned to each protein by each method.}
\end{figure}

Data were obtained using reverse-phase protein arrays (RPPA; \cite{Hennessey}) applied to a panel of breast cancer cell lines  \citep[][]{Neve}.
Data $\mathcal{D}$ comprised equilibrium observations for $p=38$ phosphorylated proteins, in addition to their unphosphoryated counterparts (Fig. \ref{VitroData}).
Cell lines belong to two biologically distinct subtypes known as {\it basal} ($n=22$) and {\it luminal} ($n=21$), with each member cell line comprising one  sample.
The true data-generating network is not known for biological samples, but for certain nodes the relevant kinase-substrate relationships have been described in considerable mechanistic detail in the literature. To minimize the risk of comparing results of inference against an incorrect literature model, we focused attention on selected nodes in the data for each of whom the key kinase is well established. 
For example, the protein S6 is known to be phosphorylated via the kinase activity of p70 S6 Kinase (p70S6K); both proteins are included in our assay. Treating S6 as the target (i.e. the network child), we scored each of the remaining 37 proteins as a candidate regulator (i.e. for inclusion in the parent set $\pi_{\text{S6}}$) using each method. 
Fig. \ref{ranks} displays the result of inference for the parents of S6 (S6 is phosphorylated on amino acid residues Serine 235,236; results for basal subtype shown; measurements of S6 phosphorylation on residues Serine 240,244 were excluded since this correlates closely with phosphorylation on Serine 235,236).
Despite the known well established regulatory role for p70S6K, 
it is striking that only our approach ranks p70S6K highly. 
The LASSO approaches ascribe no weight to the correct kinase in this case.
To gain more insight into the assignment of weights by the competing methodologies we constructed scatter plots comparing weight distributions (Fig. \ref{Vitro}). It is immediately clear that the weight assignments vary markedly between basal and luminal subtypes. 
In addition, it is noticeable that there is little agreement between the apparently similar linear formulations.
We extended this investigation to several other key signalling players whose regulation is well understood (Table \ref{table}). Overall, we find that the proposed approach outperforms the linear methods.

\section{Discussion and conclusions} \label{discussion}
In this work we investigated integration of biochemical mechanisms into network inference for steady-state data. 
We focused on protein phosphorylation, a key biochemical process where the availability of relatively sophisticated simulation models, extent of existing mechanistic insight and availability of relevant proteomic data combine to facilitate assessment of network inference approaches. 
Our results, on simulated and real data, demonstrated that protein signalling network topology may be estimated much more successfully under our approach than by conventional linear formulations. The linear approaches we used were outperformed on simulated data and failed to identify known regulation in real data.
Further, we saw that apparently similar linear formulations can return very different recommendations for which variables ought to be included in the model; one possible explanation for such disagreement may be model misspecification.
In addition to superior performance, a chemical formulation benefits from increased interpretability, ascribing mechanistic roles 
to variables and relating parameters to scientifically interpretable rates.
Although we focussed on network inference, the methodology presented here simultaneously learns values for these rate parameters, providing a fitted predictive model of signalling dynamics.

Up to this point we have not discussed identifiability of either parameters or structure. 
Parameter identifiability does not present challenges within our Bayesian formulation when the objective is structural inference. (The interested reader is referred to \cite{Craciun,Calderhead} for discussions of parameter identifiability in chemical systems.)
It terms of structural identifiability, \cite{Peters} recently 
discussed the limitations that arise from symmetry inherent in the the linear-additive-Gaussian model and showed 
that within an {\it identifiable functional model class} (IFMOC) it is possible to estimate causal relationships with statistical consistency. 
However, in order to formally show that a given functional class constitutes an IFMOC, the theory at present requires strong assumptions, including acyclicity of the causal graph, that do not hold in many real systems, including protein signalling networks.
When faced with such challenging circumstances, empirical investigations naturally have a key role to play.
In this sense, our contribution demonstrates that the theoretical results of \cite{Peters} have substantive implications for biological network inference in practice.
Further work will be required to better understand these implications and to extend the ideas presented here to further application domains, including gene regulation. 
In complementary work, \citep{Oates2} considered the use of nonlinear chemical kinetics for network inference using time-course data, reporting that a chemical formulaion outperformed a number of mechanism-free approaches, including nonparametric models.

Network inference is naturally facilitated by interventional experiments \citep{Lu}, however adequate modelling of the effects of intervention is important to ameliorate statistical confounding \citep{Pearl}.
Within a chemical kinetic framework such factors may be easily accounted for; for instance a \emph{perfect intervention} simply corresponds to removal of the targeted species from the chemical model.
In testing, not presented here, we extended our methodology to incorporate {\it imperfect certain intervention}, where the interventional targets are assumed known, but the interventions may not completely block catalytic activity of their targets \citep[see][for a general discussion of interventions in graphical models]{Eaton}. 
In the context of protein phosphorylation, kinases and their inhibitors can be intervened upon using agents (such as  monoclonal antibodies or small molecule inhibitors).
We modelled these effects by rescaling the effective concentration of interventional targets, in the presence of the treatment, as $X_j \mapsto \alpha_j X_j$ where $0\leq\alpha_j\leq 1$ is an unknown parameter capturing interventional efficacy of the agent.
Using this extended methodology we observed that interventional experiments were more informative than the global perturbation experiments considered here, leading to improved AUR scores.

Network inference based on nonlinear models is computationally challenging. 
We considered low-to-moderate dimensional settings ($p=12,38$), for which the RJMCMC proved to be  effective. 
The computations in this paper are parallelisable by population Monte Carlo techniques \citep{Laskey}, and it may therefore be possible to also extend this work to the high-dimensional setting \citep{ALee}.
In general, nonlinear approaches are clearly more burdensome than their linear counterparts, where highly efficient approaches, including those based on LASSO and related penalised likelihood schemes, allow rapid estimation even in high dimensions \citep{Meinshausen}.
We therefore view the methods presented here as complementary to  variable selection based on linear models, allowing more refined exploration in settings where some insight into underlying dynamics is available.

\begin{table}[!h]
\centering
{\begin{tabular}[b]{rcccc}
\hline
\; \; \; \; \; Target: & Akt & p70S6K & S6 & p53 \\ \hline\hline
G.K. Kinetics & {\bf 4}  & {\bf 3} & {\bf 1} & {\bf 8} \\ \hline
Lin. Bayes & 10 & 9 & 15 & 32 \\ \hline
Lin. Bayes Adj. & 14 & 8 & 8 & 14 \\ \hline
Lin. Lasso & NA & 8 & NA & NA \\ \hline
Lin. Lasso Adj. & NA & 12 & NA & NA  \\ \hline \hline
Total \# Candidates: & 36 & 37 & 36 & 37 \\ \hline
\end{tabular}}
\caption{Cancer protein data, comparison of methods. The proposed method was compared with linear approaches using reverse-phase protein array data for nodes whose regulation has been extensively studied in the literature. Each method ranked potential regulators among candidate proteins; here we display the rank assigned to the known kinase, using each of the 5 methods.
For example, Fig \ref{ranks} shows such an analysis for the target node (i.e. network child)  S6, where G.K. Kinetics ranked the known kinase p70S6K 1st out of a total of 36 candidates. 
High rank indicates that the known kinase is correctly highlighted in the analysis; the highest-ranked result is highlighted in bold for each target node.
Here we show the rank assigned to the known kinase for each of the target nodes Akt, p70S6k, S6 and p53. 
[``NA'' indicates that the known kinase received zero weight. 
Alternative phospho-forms of the target were excluded as candidates for Akt and S6, so that there were 36 candidates rather than 37.
Here we present results obtained using data from cell lines of basal subtype; luminal results are shown in Supplementary Information.]}
\label{table}
\end{table}

On real proteomic data we observed that network inference was challenging.
In particular, inference based on data obtained from luminal cell lines encouraged poor performance from all approaches (see Supplemental Information).
Whilst the genomic background (in our example breast cancer) may be a factor - illustrated by the luminal failure case - we suspect that the real difficulties result from a complex noise process.
At present, network inference can aid in hypothesis generation, but care must be taken in interpreting results.
Further experimental and methodological advances will be required before network inference methods can be regarded as  truly robust tools for biological discovery.

In this work we investigated integration of biochemical mechanisms into network inference. 
Whilst the Goldbeter-Koshland formulae are invalid at the single-cell level, which is intrinsically stochastic, the evidence presented here suggests that these deterministic nonlinear equations represent a better approximation than the corresponding linear equations.
In particular a chemical kinetic formulation is able to account, in a principled way, for variation in total protein levels between samples.
Consequently, inferred edges cannot be interpreted as indicators of direct biochemical interaction; rather an edge corresponds to the prediction that intervention on the parent will result in a change in expression of the child, possibly indirectly via unobserved variables.
In our real data example we therefore allowed for candidate species which are not themselves kinases, such as S6 and p53.

For simplicity, in specifying the class $\mathcal{F}$ of functional forms, we did not consider post-translational modifications such as ubiquitinylation, nor spatial effects such as translocation, nor did we explicitly distinguish between phosphorylation on different residues. 
The methodology which we presented may be generalised to other molecular mechanisms. In particular alternative mechanisms of enzyme interaction such as noncompetitive, uncompetitive, hyperbolic and parabolic inhibition could be readily integrated into our framework.

\section*{Acknowledgement}

\paragraph{Funding\textcolon} 
Financial support was provided by NCI CCSG support grant CA016672, 
NIH U54 CA112970, UK EPSRC EP/E501311/1 and the Cancer Systems Biology Center grant from the Netherlands Organisation for Scientific Research.

\end{document}